\documentclass[twocolumn,showpacs,floatfix]{revtex4} 
\usepackage{graphicx} 
\usepackage{dcolumn} 
\usepackage{amsmath} 
\usepackage[latin1]{inputenc} 
\begin{document} 
\title[How a spin glass remembers.]{How   a spin-glass remembers. Memory and rejuvenation
from intermittency data: an analysis of temperature shifts. }  
\author{$^{1,2}$Paolo Sibani\footnote{Permanent address: Fysisk Institut, SDU,  Odense, DK} 
 and $^{2}$Henrik Jeldtoft  Jensen} 
\affiliation{$^1$Theoretical Physics, Oxford University, 1 Keble Rd, Oxford OX1 3NP, UK\\   
$^2$Imperial College London, South Kensigton Campus,   SW7 2AZ, UK} 
\date{\today} 
\begin{abstract}  
The memory  and rejuvenation  
properties  of intermittent 
heat transport   are  explored theoretically
and by numerical simulation for  Ising spin glasses with  short-ranged
interactions. 
The theoretical part develops 
a  picture of non-equilibrium 
glassy dynamics recently introduced by the authors, 
which  links irreversible `intermittent'
events, or `quakes'  to thermal  fluctuations of record magnitude. 
The pivotal idea  is that   the 
 largest energy barrier $b(t_w,T)$ surmounted 
 prior to $t_w$  at temperature $T$  
 determines  the rate $r_q \propto 1/t_w$ of the 
 intermittent events   near $t_w$.
The idea  implies that the same rate 
 after a negative temperature shift 
should be  given by $r_q \propto 1/t_w^{eff}$. The 
  `effective age' $t_w^{eff} \geq t_w$
has an algebraic dependence on $t_w$, whose
 exponent contains   the  temperatures
before and after the shift. 
This  analytical expression is confirmed  by  numerical simulations. 
Marginal stability 
suggests an asymmetry between cooling and heating,
i.e.\  a  positive temperature  shift $T \rightarrow T'$ 
should  erase  the  memory   $b(t_w,T)$.
This is  confirmed by  the simulations, which 
show a rate $r_q \propto 1/t_w$, controlled by   
the barrier $b(t_w,T') \geq b(t_w,T)$.  
   Additional 
`rejuvenation'  effects
are also  identified  in the intermittency data 
for shifts of both  signs. 
 \end{abstract}
\pacs{65.60.+a;05.40.-a;75.10.Nr} 
\maketitle
 \section{Introduction} 
   Many fascinating aspects of non-equilibrium
   glassy dynamics are revealed  by 
  temperature  changes  applied  while the system ages, both 
    experimentally, see e.g.~\cite{Granberg90,Lederman91,Jonason98,Jonsson02}  and 
    in numerical simulations~\cite{Komori00,Berthier02,Takayama02}.
Spin-glass experiments  often measure  the out of phase 
AC susceptibility~\cite{Jonason98}$\chi''(\omega,t_w)$,  or the 
(related) derivative of the ZFC magnetization
with respect to the logarithm of time $S(t,t_w)$. In the protocol
introduced in ref.~\cite{Jonason98} 
the temperature is decreased at a constant  rate, except for one 
temperature  $T_0$ which is maintained  for e.g. hours.
During this aging stage, $\chi''$ decreases, revealing the characteristic 
dynamical stiffening  generally  present in glassy systems.
As  cooling is resumed, the susceptibility  returns to the $T$ dependence 
of    an un-aged system. The time spent aging at  one temperature
is thus  of little avail  at  a lower temperature, and  the $\chi"$ versus $T$ curve
features a dip  centered at $T_0$. 
That this  dip is reproduced when re-heating  at a constant rate through $T_0$ 
 demonstrates  the persistence of the memory of the configurations   visited.
A later extension of this work~\cite{Jonason00} shows  that 
two  dips can be produced and retrieved by aging at 
different temperatures. More importantly, the memory can be erased 
by overheating, i.e.\ in a second   cooling  phases  the dip is no
longer present,
provided that the intervening temperature has been sufficiently high.
 
In  temperature shifts experiments, the system is aged for a time $t_w$,
 and a small
magnetic field is then turned  on  to probe the dynamics.  Under   isothermal 
conditions,  $S(t.t_w)$  has a maximum at $t=t_w$. If a small temperature
shift of either sign is   imposed at $t_w$, this maximum moves~\cite{Granberg90}.
The  position of the maximum  on the $t$ axis defines 
 an  effective age, which is   studied 
as a function of  various parameters.
Ref.~\cite{Jonsson02} is particularly concerned with 
the issue of the invertibility of the relation between 
age and effective age, an issue which 
we return to in the discussion. Numerical simulations
are also able to follow the time dependence of the energy density
$e(t)$ after a $T$ shift: the so called 'Kovacs' effect is best observed
when  shifting the temperature from $T_1$ to $T_2$ at a late time,   
for which the energy is already close to its equilibrium value at $T_2$: 
After the shift,  $e(t)$ approaches  the isothermal energy  density 
  at  $T_2$ in a non-monotonic way. This effect can be used to define 
an effective age~\cite{Komori00,Berthier02} as a time shift allowing
one to superimpose the the $T$ shifted $e(t)$ onto its isothermal
counterpart at the final temperature.  
  
 The present investigations   give only indirect insight  on configurational
changes and how these changes are memorized by the aging spin-glass.
They do, however directly shed light on   a related issue, 
i.e. how the size of the dynamical barriers surmounted in an aging system
is memorized. The data
   allow a parameter-free theoretical interpretation  by  a recent   model
    of non-equilibrium  
     relaxation~\cite{Sibani03}, which provides a unified
     description of pure aging by   attributing 
     a pivotal role  to   energy fluctuations of record size.
The simulations  combine  temperature shift techniques
  with a statistical analysis
    of heat transfer~\cite{Sibani04} motivated by the intermittency    
    studies  of refs.~\cite{Cipelletti03a,Buisson03,Buisson03a} in soft-condensed
matter systems and glasses. 
In this way,  many important aspects of aging 
dynamics can be extracted from a one-time quantity, the energy. This  is   potentially  useful 
      whenever  eliciting a linear response is 
     difficult or impossible. 
 \begin{figure}[ht!]
\includegraphics[width=8cm]{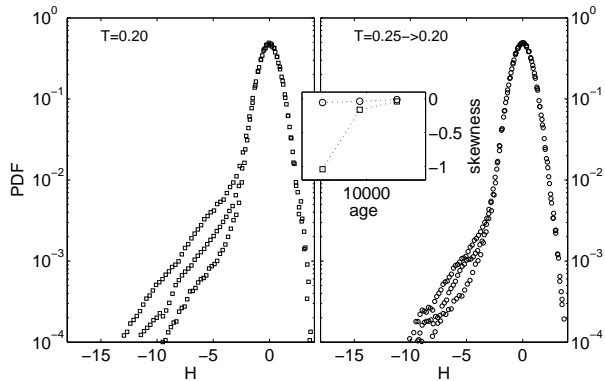}  
\caption{ 
Left figure (squares):  the PDF's  
 of the energy differences
$H(\delta t, t, t_w)  = E(t_w+(n+1) \delta t) -E(t_w+n \delta t)$,
$n=0,1,2 \ldots$, $n\delta t \leq t$ for isothermal aging at $T=0.20$. 
The  three  data sets are taken in the 
observation  windows  $t_w,t_w+t$, with 
$t_w= 4400, 8800$ and $17600$. The observation time and 
sampling interval are    $t=1200$ 
and  $\delta t=22$. 
 The PDF  is seen to approach  a Gaussian 
as the age increases while  the other parameters remain fixed.
Right figure(circles): all parameters are as above, but  
the system is now  initially quenched to $T=0.25$ and 
further cooled to $T=0.20$ at $t_w = 2140$.
Insert: the skewness of the PDFs as a function
of the age at which the data collection commences.
Squares and circles
correspond to isothermal and temperature shift conditions.  
In the latter 
case the skewness has a considerably 
smaller variation with the age, i.e.\ the  PDFs differ less among
themselves than the isothermal ones. The numerical 
values are  smaller  and similar to those
of  older isothermal system. 
} 
\label{introplot}  
\end{figure}  
The definitions  used
for the  effective age  are  based on the 
aging  behavior of the rate of intermittent events,
and  differ   from the definitions  used in  
studies of  linear response effects and energy density,
as discussed in detail at the end of the paper. 
Nevertheless, whenever a comparison
is possible, concurring  physical conclusion can be  drawn.

The Probability Density Function (PDF) of the heat flow data shows that two 
 types of  events occur: reversible fluctuations, 
which are described by the Gaussian part of the  PDF, and
large,  irreversible events, dubbed quakes, through which the excess
energy initially trapped in the sample is transferred to the heat bath.
These events appear as an asymmetric exponential tail of the PDF.
The general  form of the PDF is easily understood:  thermal fluctuations
occur independently and simultaneously in different part of the systems,
whence their sum is Gaussianly distributed. Occasionally, a thermal 
fluctuation induces  a large  configurational rearrangement
associated with  a correspondingly large energy release.
Such  a quake 
is arguably  composed of a series of consecutive 'downhill'  
 flips of contiguous spins, and its size is exponentially distributed if each movement 
can be the last one with equal probability.

A closer inspection of the data reveals that 
the rate $r_q(t_w)$ of the quakes: 
  \emph{i)} is temperature
independent~\cite{Dall03,Sibani04}, which 
is striking for activated dynamics,  and \emph{ii)}: decreases, under isothermal conditions,
as the reciprocal of the  age $1/t_w$. 
The effective age $t_w^{eff}$  of a sample, possibly  subject to $T$ shifts, 
can hence be   defined  as  $1/r_q$, and  its behavior 
can be empirically investigated under different conditions.
   The    theoretical idea  used to explain these  findings 
 is  that   a glassy system  
    `remembers' the largest or `record' energy  barrier $b(t_w,T)$  
    surmounted  during its (iso)thermal history 
    up to time $t_w$~\cite{Sibani03,Sibani04}. Specifically, 
    only    fluctuations  exceeding $b$  are able to produce 
the   intermittent events on each time scale.
    For isothermal  aging,  record dynamics  explains  the 
 decelerating rate of events    $r_q(t_w) \propto  1/t_w$ and   accounts for  the
temperature independence of  $r_q$. 
 Within this framework,   a mathematically consistent explanation 
is only possible   if the dynamically available 
configurations are distributed in energy 
according to   an exponential `local density of states' (LDOS), a 
 form which is   widely observed in complex systems. 

 In this paper, we draw a further consequence  of these 
ideas, by predicting     the effect
  of a   temperature shift  imposed at $t_w$ on  
  the   rate $r_q(t_w,T,T')$ of intermittent events.  
Unlike the isothermal case, 
 this rate could  depend, beside  $t_w$ itself, 
on the temperatures $T$ and $T'$ before  and after the shift.
For  negative temperature  shifts, the   form of $t_w^{eff}$ is 
derived analytically under the assumption that 
the memory of the extremal barrier $b(t_w,T)$ established prior to the shift
persists and that its value 
 still determines the evolution at the lower temperature $T'$.

A positive   shift allows stronger energy fluctuations which
very quickly   overcome $b(t_w,T)$ and reach 
  $b(t_w,T')$,  the extremal barrier 
for  isothermal relaxation  at the final temperature $T'$.
Hence,    the intermittent energy release  continues
at a rate  $r_q \propto 1/t_w$.
These properties are  fully confirmed by the simulations.   
  
A second observation is that 
 the extremal barrier does not 
 account for  the proportionality
constant between $r_q$  and   $1/t_w^{eff}$ 
(or $1/t_w$).  This constant  is larger (i.e.\ stronger intermittency)
after   shifts of either sign  than for  the isothermal relaxation. In  this
sense,  the shifts 
make the system appear younger, and induce 
what could be called a rejuvenation effect.
  
    In the next section  we summarize the  theoretical
    ideas linking non-equilibrium drift to record  fluctuations.
In section~\ref{theory} we derive the formula
for the effective age, which is tested in Section~\ref{results},
where  the numerical
    results are also given.  
    Section~\ref{technical}  provides technical details
    on  the simulations and on the   data handling.
       and section~\ref{discussion} is a  brief summary 
    followed by a discussion of  how the present
results relate to  previous investigations. 
	      
 \section{Why record  fluctuations could be 
important in glassy systems}
 \label{theoryback}
Extremal fluctuations  are largely  irrelevant
 in   equilibrium,  but may have pivotal
effects in metastable situations, i.e.\ 
 when  exceeding a threshold can push  the system 
into a completely different dynamical path. 
 Consider first,  as an example, 
  an artifact, e.g.\ a skyscraper,   
which   collapses  if hit by  seismic events 
above  a certain  magnitude.  
We assume that,  against good  engineering
 practice,    
a `marginally better' skyscraper is  
built following  any  collapse,   which is able to    resist  
    events  up to the size 
  causing   the  latest break-down, but unable to resist 
any---even   slightly---stronger event. 
In the   sequence  of 
gradually stronger  skyscrapers   built  according to
this prescription, 
each  element  collapses  
if   hit by an earthquake stronger
than all its predecessors. The 
  magnitude is irrelevant, and the 
statistics of    record-sized earthquakes
 describes  in this case    physical
events. 

Consider now a sequence of  random numbers
drawn independently from the same 
continuous distribution.
The distribution of the  number $n$ of records, i.e. numbers
larger than all previous extractions,  
is  independent of the distribution from which 
the data are extracted~\cite{Sibani93a}. 
An approximate  statistical  description of the 
records~\cite{Sibani93a,Sibani03} 
 is the log-Poisson distribution 
\begin{equation}
P_n(t_1,t_2) =  \frac{(\alpha \log(t_2/t_1))^n}{n!} (t_2/t_1)^{-\alpha},
 \quad \alpha \leq  1, \quad t_2 \geq t_1  
\label{logP}
\end{equation}
where $P_n$ is the probability 
that $n$ records will occur 
between trials  $t_1$, 
and where the parameter $\alpha$  
describes that the records can 
occur in  $\alpha$ independent  processes 
running  `in parallel'.
We shall later think of these    trials  as a time  
sequence, and of the random numbers  
as energy values produced by thermal fluctuations.
Furthermore we will mainly need the   formula
in the limit where $t_2 \approx t_1 $. Glossing over the
discreteness of these parameters, and with $t_2=t_w$ 
the  rate  of   events near  $t_w$   falls off as $r_q \propto 1/t_w$.

Marginal stability provides a physical mechanism by which 
the rank, and not the magnitude of the fluctuations
determines  the non-equilibrium dynamics associated with 
changes of attractor. Because the temporal statistics of record events
is system independent, the concept is potentially  useful  whenever  metastability is 
important.  Its    physical  relevance   
was first noticed in a simple model  
 of Charge Density Waves dynamics~\cite{Tang87},   later 
extended to describe the glassy dynamics of  various
systems with multiple metastable
 configurations~\cite{Sibani93a,Sibani95,Sibani99},
and more recently 
 to thermal aging~\cite{Sibani03,Dall03,Sibani04} of glasses.
The numerical investigations of ref.~\cite{Dall03} 
show explicitly how selected  energy records characterize 
the attractors visited during unperturbed thermal aging after
a quench.  

\emph{A priori}, marginal stability  can  
play a dynamical r\^{o}le whenever: 
i) Numerous  attractors, or `intrinsic states'
exists , with different degrees  of  stability, the latter gauged by
a typical exit time  or, equivalently, exit energy  barrier. 
 A \emph{marginal} increase
 of   the   degree of  stability  of the attractor
  selected can arise from 
an entropic effect, i.e.\ simply reflect   
an overwhelming predominance of  shallow attractors. 
 
ii) the initial deep quench 
typically produces configurations with a large
energy excess. Similar to   the  gravitational
energy of a  building    released irreversibly during
its collapse, this  energy
can be  released in relative large and (hence) irreversible
bursts~\cite{Sibani03,Sibani04}.

As in   the  skyscraper example, if  the increase in 
attractor stability is marginal,  
the non-equilibrium dynamics
is    driven by fluctuation extrema. The 
 number of quakes  occurring in the interval $[1,t_w]$ is 
then described by 
the  log-Poisson distribution  given in Eq.~\ref{logP}, with
$t_1=1$ and $t_2=t_w$. 
 
To reach     quantitative predictions,   some   
basic aspects of real-space morphology 
 must  also  be considered: 
Aging  in spin-glasses with short-ranged interactions
is characterized   
by a  slowly  growing length scale,  
the thermal correlation length,   
which  remains  
 shorter than a few lattice spacings~\cite{Kisker96,Berthier02}
for accessible time scales.  
It is therefore reasonable to assume that in  an extended 
 system with short range  interactions  many 
thermalized sub-domains  
 exist, whose linear size matches 
 the thermal correlation length and which 
  are separated by frozen degrees of freedom. 
   Their   reversible thermal fluctuations are  statistically
  independent and are described,    
through   the usual tools of statistical
mechanics,   by    a `local' energy landscape comprising, 
for each domain, the configurations available.

The   model   assumption  explored in 
the following is  that 
 an extremal fluctuation 
 within each sub-domain is able to
trigger a quake~\cite{Sibani04,Sibani03}. 
This has two consequences: Firstly, the total number of quakes
 is  a sum of independent Poisson  processes, and hence
itself   a Poisson process with average $\alpha>1$.
This average   increases  with the  system size and  
 attains the smallest possible value   $\alpha=1$ 
only if    a single sub-domain is  present.  
Secondly, as we argue below, combining the 
thermal nature of the  extremal fluctuations 
with the properties of record statistics    
means that the sub-domains must have 
a nearly  exponential local density 
of states LDOS, of the type
   observed numerically in a variety
of small glassy  systems~\cite{Sibani93,Sibani99,Schon00}.
 
Ignoring for simplicity any difference
between the sub-domains, we   write this  LDOS  as 
 \begin{equation}
      {\cal D}(\epsilon)\propto \exp(\epsilon/\epsilon_0), 
      \label{expD}
      \end{equation} 
 and note that the  condition $T <\epsilon_0 $ is required  in order to 
  ensure the  thermal 
metastability of any attractor to which this LDOS applies.

Importantly,   a log-Poisson description is only possible
as long as a   sharp distinction between
 local reversible fluctuations and irreversible changes is
 meaningful. The limitation does not seem    unduly
restrictive in  deeply quenched 
 glassy system, which for many decades of time  remain 
 far from  true thermal  equilibrium.  
The  statistical 
 independence of fluctuations
which is required by  the theory,  is approximately
 fulfilled at low temperature, 
as trajectories  mostly dwell   at the bottom
 of local wells. 
  
The above  ideas can  be  directly tested in  intermittency 
measurements, where 
different parts of the same data set 
 provide the required information~\cite{Sibani04}. 
 E.g.\ the left 
panel of Fig.~\ref{introplot} shows the
Probability Density Function (PDF) 
of the heat transferred over a small time interval $\delta t$
under isothermal conditions. 
  \begin{figure}[ht!] 
\includegraphics[width=8cm]{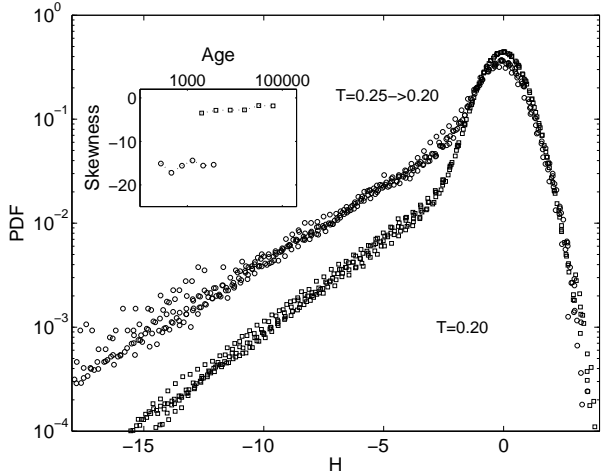}
\caption{ 
Main figure, lower data sets:  the PDF's  
  of the energy differences
$H(\delta t, t, t_w)  = E(t_w+(n+1) \delta t) -E(t_w+n \delta t)$,
$n=0,1,2 \ldots$. The five sets correspond to ages
$t_w=100 \delta t$ with 
$\delta t = 10,20,40 \ldots 160$  and 
observation window $t=t_w/2$. The temperature is $T=0.20$.
Upper data sets:   
One additional  cooling step  takes the system from   $T=0.25$ to 
$T=0.20$.  For the $i$'th data set, $i=1,2, \dots 5$ 
the change occurs at age  $t_{wi}$, determined such that 
  the corresponding 
\emph{effective} ages  are 
   $t_w^{eff}= 2000, 4000, 8000, 16000$ and  $32000$. The
corresponding sampling intervals are 
$\delta t =t_w^{eff}/100$ and $t=t_w^{eff}/2$.  
Keeping the ratio  $\delta t/t_w^{eff}$ constant collapses the  data,
whence the effective age plays the same role as the
actual age does in the isothermal case.
The insert shows that the skewness of the PDF's 
 falls into two distinct groups, with only a small variation
within each group. 
 } 
\label{negstep} 
\end{figure}
The   Gaussian part of the PDF describes the
reversible fluctuations and is found to be  
 independent
of $t_w$~\cite{Sibani04}, indicating that 
the attractors successively explored 
are similar   with respect to their equilibrium energy
fluctuations, as already implied by Eq.~\ref{expD}, even though 
 they may differ with respect to their   
 stability against fluctuations. 
The temperature dependence of the variance  
of the  Gaussian fluctuations  
agrees  with the  equilibrium value
calculated from 
Eq.~\ref{expD}, lending quantitative  support 
to this assumption~\cite{Sibani04}. 

The exponential tail describes the intermittent
events, which   are rare but 
far more frequent than Gaussian fluctuations of 
comparable size. The tail  has a   clear  
age dependence, which is  visible in  
 Fig.~\ref{introplot}, and which gets stronger 
the smaller $t_w$ is.   An increase 
of the weight of the tail is  hence tantamount to 
a rejuvenation effect.  The rate  
of intermittent events falls  off as $r_q \propto 1/t_w$, 
as demonstrated by the collapse 
of   a set of PDF's obtained for  different
$t_w$ and $\delta t$ with constant   ratio $\delta t/t_w$.
In what follows,  we use the  collapse  
of heat transfer PDF to draw inferences  on the 
rate of intermittent events, and hence on the 
`effective age' of the system, after 
a small temperature shift.  
 \section{How a spin-glass remembers}
\label{theory}
If  uncorrelated  energy records trigger the  
quakes, the  time  intervals between these quakes, 
i.e. the residence time $t_r$  spent in  
metastable states, grows  in proportion to the age~\cite{Sibani03}
and is independent of the temperature.
This  so-called `pure scaling'  behavior is well known 
and    analytically predicted 
for   log-Poisson statistics. 
We presently choose to ignore    the  usually small but
always  systematic 
deviations from  pure aging, sub- or super-aging,  which appear
 in  both numerical and experimental results.

\begin{figure}
\includegraphics[width=8cm]{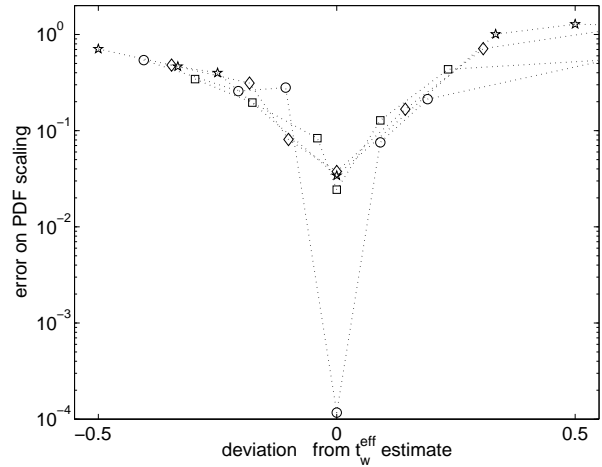} 
\caption{ \small  
The figure shows  how the 
quality of the data collapse, of the sort
shown in Fig.~\ref{negstep},   deteriorates as
the condition of constant  $\delta t/t_w^{eff}$ is violated. 
The effect is achieved by changing the value of $\delta t$, which    
amounts  to   varying  $t_w^{eff}$ around its theoretical
value given  by Eq.~\ref{effective}. The abscissa is the
relative deviation, and zero thus corresponds to the theoretical
prediction.     
   The following  negative $T$ jumps are considered 
$0.25\rightarrow 0.20$ (circles), $0.35\rightarrow 0.30$ (squares), 
$0.45\rightarrow 0.40$ (diamonds)and  $0.55\rightarrow 0.50$ (stars).
Lines are  guides to the eye.
Note that in all cases the best collapse is obtained at or very
near the predicted  value of the effective age. 
 }
\label{scaling_error} 
\end{figure} 
A second expression for $t_r$ can be 
derived using that   the system is thermally 
equilibrated in the metastable states
between  the quakes.   In this case,   hopping over
the  extremal barrier $b(t_w,T)$ (see below) 
is mediated by  \emph{equilibrium}  thermal fluctuations.
Hence, the residence time in a metastable
state entered on a time scale $t_w$ is proportional to 
the   Arrhenius  scale $t_r = \tau_A \propto \exp(b(t_w,T)/T)$.
Consistency with our previous
argument requires   $\tau_A \approx t_w$ which is  mathematically 
possible  only   if   
the  LDOS from which the 
energy fluctuations are drawn has the exponential form 
given by   Eq.~\ref{expD}. 
     	
  Equation~\ref{expD} describes the distribution in energy 
      of the configurations   accessible to 
    an arbitrary   thermalized sub-domain, with   zero 
     defined as the  lowest  available energy.  
     The  probability of visiting states   of energy
larger than $\epsilon$ without exiting the attractor    is 
then 
\begin{equation}
P_{eq}(\epsilon) = \exp(-\epsilon a(T)),
\label{Boltz}
\end{equation} 
where the reduced inverse temperature 
\begin{equation}
  a(T) = (1/T -1/\epsilon_0)
\label{reduced} 
\end{equation}
describes, for temperatures well below $\epsilon_0$,
 the entropic effect 
of the exponential divergence of the LDOS.
 Throughout, the value
$\epsilon_0 = 0.86$ is used. This value, which is
close to the critical temperature of the \emph{3d} 
Edwards-Anderson spin-glass~\cite{Bhatt88}, 
was obtained in ref.~\cite{Sibani04} by fitting 
 the theoretical
form predicted by Eq.~\ref{Boltz}
to the   heat capacity data inferred from the Gaussian
heat fluctuations.
 
The largest fluctuation  occurring  during   $(0,t_w]$
is the largest of  ${\cal O}(t_w)$ energy values
 drawn independently from the exponential equilibrium
distribution~(\ref{Boltz}). According to a standard result
of extremal statistics~\cite{Leadbetter83} (NB:
this  result only requires that the   
tail of the Boltzmann distribution be  exponentially
distributed)   its   average value grows with the age as  
\begin{equation}
 b(t_w,T)  =  \ln(t_w)/ a(T) +{\rm const.}, 
\label{max}
\end{equation}
whence the    Arrhenius time $\tau_A = \exp(a(T) b(t_w,T))$
remains equal to $t_w$, as anticipated.

The above argument 
is now   extended   to the situation 
 where the temperature is instantaneously lowered from $T$ to
$T'$ at age $t_w$:  the   
 extremal barrier $b(t_w,T)$ previously established
remains the same, but must 
now be overcome  at 
a lower temperature in order to induce a quake.	 
  The
  relevant  Arrhenius  time  is  then  given by  
\begin{equation}
\tau_A = t_w^{eff} \propto  \exp(a(T') b(t_w,T)) 
 \propto t_w^{\frac{a(T')}{a(T)}}
 \quad T' \leq T ,
\label{effective}
\end{equation}
where Eq.~\ref{max} is 
used for  the right-most equality.
The   rate of intermittent events after
the shift  is    $r_q \propto 1/t_w^{eff}$.
Note that the  argument ignores    the decay of the rate 
occurring  while the statistics is
collected. 
In the cases considered, the   drift remains  small 
since  the  observation time is well below the 
effective age.

Equation~\ref{effective} implies  that  the PDF's 
of the  heat transfer obtained for various $\delta t$ and $t_w$
values   collapse 
if the  ratio  $\delta t/t_w^{eff}$ is kept  constant. This 
is demonstrated in Figs.~\ref{negstep}~and~\ref{scaling_error},
which  show that  
$t_w^{eff}$  plays   a   r\^{o}le  
 analogous to   $t_w $  in isothermal aging experiments 
(See Figs.~\ref{negstep}~and~\ref{posstep_n} and ref.~\cite{Sibani04}).
As    anticipated, the step has an  additional   effect:  
 the  collapse leads to a   master curve  
with a considerably stronger exponential tail
than the  curve  correspondingly 
obtained from  isothermal data.
  
In summary, a temperature drop has
two effects: \emph{a:}\  It effectively increases
the size of all  barriers, leading to an effective 
age greater than the age. In this respect the 
system looks older. \emph{b:}\  The amount of 
heat delivered by the intermittent events is higher,
 whence the system looks  in this respect younger. 
As later discussed, a similar  duality 
is seen  in refs.~\cite{Berthier02,Takayama02}. 
A positive temperature shift   
 qualitatively differs 
from a negative shift. The  
 extremal barrier $b(t_w,T)$ still
sets the time scale for the \emph{first} intermittent event  after
the step. Since  this barrier  is now strongly  
reduced  relative to the typical size of the  thermal  
fluctuations,  its  crossing is   neither    rare 
nor   necessarily  irreversible. The  additional time  
needed to reach  the extremal barrier $b(t_w,T')$ 
is, clearly, bounded by  $t_w$, whence  we can  expect the data
to be scaled by keeping  $\delta t/t_w$ constant.
This is confirmed by  
 Fig.~\ref{posstep_n} showing  an excellent
data collapse after the  positive shift.
Again, comparing the master curve  with  its isothermal counterpart
shows a stronger exponential tail. Hence, a  positive step 
rejuvenates the dynamics in the same sense 
as   discussed in item \emph{b} above.  

 \section{Simulation results}
\label{results}
The simulation results  
are now described in more  detail. 
 As in  ref.~\cite{Sibani04},  the  
	 statistics of the amount  of heat exchanged 
         over  a  time interval  
	 of length $\delta t < t_w$ is collected for
	 $n$ contiguous  intervals, with $n$ chosen such
	 that   $n \delta t = t$. The $k$'th value sampled is thus
\begin{equation}
H_k = E(t_w +(k+1)\delta t)-E(t_w+k\delta t), \quad  k \leq n,
\label{heat_def}
\end{equation}
where $E$ is the  energy of the system.
Since  the  statistical properties of the heat transfer
depend on $\delta t$, $t$ and $t_w$ only,  the process itself is
 denoted by  $H(\delta t,t,t_w)$. 
	  The drift toward lower energies is  described by  
	 an asymmetric exponential tail, and the fluctuations
	 by the Gaussian part of the PDF.
         For  fixed  $t$ and $\delta t$,
         the   PDF's  approach 
         a Gaussian  shape  as $t_w$ increases and   
	 the  skewness (third central moment) correspondingly approaches zero from below.  
         This is  seen   in the left frame of Fig.~\ref{introplot},
         and in  the corresponding  insert. The three 
	 PDF's   are taken at $t_w=4400, 8800$ and $17600$,
	 all with with $t=1200$, $\delta t=22$ and $T=0.20$. 
         The data shown in the right hand panel of Fig.~\ref{introplot}  
         are obtained with  the same parameters, except that the temperature
         is instantaneously decreased from $T=0.25$ to $T=0.20$ at 
	 $t_w = 2140$. The Gaussian fluctuations are not affected by the step.
         The tails, however, are less pronounced than in the isothermal case,
         and thus qualitatively similar to those of an isothermally aged
         but older system.

Figure~\ref{negstep} contains twelve data sets,  
           which fall into two distinct groups of equal size.
	 Data with the most pronounced  intermittent tails 
        are taken after  instantaneously decreasing  
         the  temperature   from $T=0.25$ to $0.20$ 
        at ages $t_w=275, 460, 770, 1283,2143$ and $3578$. 
        According to   Eq.~\ref{effective},  
         these values  correspond to the
       effective ages $t_w^{eff}=2000,4000,\ldots 64000$.
          The data collection starts immediately after the 
         step and continues in each case up to the age
        corresponding to  $(3/2)  t_w^{eff}$. 
        The energy differences in the r.h.s. of Eq.~\ref{heat_def} are taken
        over   a time interval  $\delta = t_w^{eff}/100$. 
        For each set    of physical parameters, $6000$ independent runs were
        performed.  The   collapse  demonstrates that the rate of intermittent events 
        is proportional to the theoretical prediction $1/t_w^{eff}$. 
        For comparison,  six data sets were  collected    isothermally, 
         from age $t_w$ to age $(3/2) t_w$,  with  $\delta t= t_w/100$
        and $t_w= 2000,4000,8000,\ldots 64000$.
        They collapse among themselves 
        but, even though the $t_w$ values
        pairwise correspond   to the $t_w^{eff}$ values for the temperature step,
        the two groups of data remain clearly distinct.
        This is the dual effect previously mentioned. 

  \begin{figure}[ht!]
 \includegraphics[width=8cm]{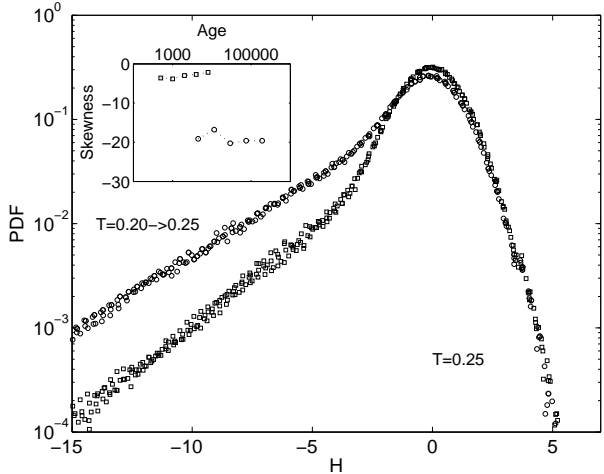} 
\caption{ \small  
Main figure,  upper  data sets:
the  heat transfer PDF's  collected after a temperature  step
 imposed at $t_w= 1000,2000,4000,8000$ and $16000$,
 i.e.\ in the interval  $t_w$ to $(3/2)t_w$. The  energy differences
are taken over $\delta t = t_w/100$.  
Keeping $\delta t/t_w$ constant is seen to  
achieve  an excellent  collapse of the data.
Lower   data sets: the same quantities  are obtained under
isothermal conditions at $T=0.25$.
The insert shows that the skewness values of the data sets
fall into two clearly distinct groups.     
  }
\label{posstep_n}
\end{figure} 

        Very similar results  were  obtained for  
        other  temperature steps , i.e.\ $0.55\rightarrow 0.50$,
        $0.45\rightarrow 0.40$  $0.35\rightarrow 0.30$.
        Figure~\ref{scaling_error} 
        shows that the quality of the results is  highly 
        sensitive to the value chosen for $t_w^{eff}$, with the
        best results obtained at or near the theoretical value. 
	The ordinate  empirically  quantifies   the  quality of the data collapse 
         of two PDF's,  say $a$  and $b$, and is obtained  
        as follows. 
       First, the extreme tails, which are mostly  affected by
       statistical error, are cut off. We then let $p(a)$ and $p(b)$ denote   the
       probabilities carried by the rest of the tails.  
        The   error   is defined as
      $e(a,b) =  \mid (p(a)-p(b))\mid  (p(a) p(b))^{-1/2}$. 
      For each data set,   $p$ is  calculated over an interval  
      extending  over $8$ units of energy, i.e. from $-4$ to $-12$ 
      for the $0.25\rightarrow 0.20$ data. This interval
      is gradually  moved to the left
      (more negative values) for $0.35\rightarrow 0.25$ etc., 
      to match the increasing width of the  Gaussian part of the
      curves as the temperature  increases.

The upper data set 
in Fig.~\ref{posstep_n} shows the  data collapse 
obtained after  a positive temperature shift. 
As was the case for the negative shift, the  
Gaussian parts  are  unaffected, while the    
intermittent tails are strongly 
affected. 
The  set, comprises six different
PDF's from systems of different ages,
$t_w=2000,4000,\ldots 32000$,
with the ratio $\delta t/t_w=1/100$. 
Data are collected from $t_w$ to $(3/2)t_w$,
and $2000$ independent runs were
performed for each set    of physical parameters.   Since, as the     data
collapse  demonstrates,   the rate of intermittent events 
remains  proportional to  $1/t_w$,  the stronger  fluctuations
available  after the  shift are able
to   quickly overcome all  barriers
in the range between  $b(t_w,T)$  and $b(t_w,T')$,
the extremal barrier  which
would have been reached isothermally at the higher
temperature. The latter  barrier hence   sets the rate for
the intermittent events.  
The lower data sets comprise   five  isothermal PDF's
obtained under the same conditions,  but for fixed temperature $T=0.25$.  
The  presence of two distinct master curves
shows, again, the presence of
an additional rejuvenation aspect.   

\begin{figure} 
 \includegraphics[width=8cm]{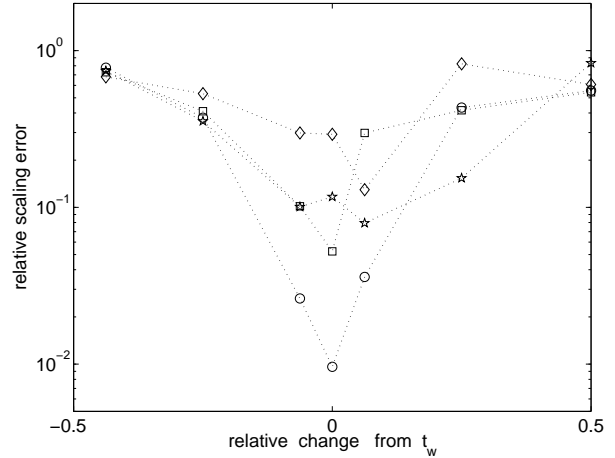} 
\caption{ \small  
\small   The figure explores how the 
quality of the PDF  collapse after a positive
temperature step---of the sort   shown 
in Fig.~\ref{posstep_n}---deteriorates as
the criterion of constant
$\delta t/t_w$ is systematically violated.  
The abscissa is the relative deviation of the 
scaling parameter used from $t_w$. 
   The following  positive  $T$ jumps are considered 
$0.20\rightarrow 0.25$ (circles), $0.30\rightarrow 0.35$ (squares), 
$0.40\rightarrow 0.45$ (diamonds) and  $0.50\rightarrow 0.55$ (stars).
Lines are guides to the eye.
}
\label{scaling_error2}  
\end{figure}

Finally, Fig.~\ref{scaling_error2} explores how the quality 
of the   collapse of PDF's obtained after a
positive temperature declines, as the criterion $\delta t/t_w=$constant
is  systematically  violated. These  results, obtained for 
initial temperatures $T=0.20,0.30 \ldots 0.50$, 
upward shifted  in each case by  $0.05$,   show that the conclusion
drawn from Fig.~\ref{posstep_n} remains valid for a
range of  temperatures.  

 \section{Technical details}
    \label{technical} 
        All simulations are performed for   a $16^3$ 
	Edwards-Anderson  spin-glass  			   		       
        model with nearest neighbor Gaussian couplings~\cite{Edwards75}
	of unit variance and zero mean. The  Boltzmann constant is 
	set to one.  
	 
         We employ  an  event driven simulation technique, 
	the  Waiting Time Method (WTM),  also used in  
	the landscape exploration~\cite{Dall03}. 
        The   `intrinsic' or current   time  of the system is initially
	zero and corresponds  for large system sizes  to 
        the usual  Monte Carlo sweep. 
	A  slight  modification of    the WTM procedure
	described in~\cite{Dall01} is required to deal 
	with temperature  changes    beyond the initial 
	instantaneous quench at $t_w=0$,  
	if the mathematical equivalence of the WTM 
	to the standard Metropolis
	algorithm is to be maintained. This is discussed below 
        together with   
	the main points of the algorithm. 
	 
	Let $\delta E_i$ be  the energy difference  
	produced  by  flipping  the $i$'th spin. Initially,    
	random waiting times   are drawn independently from   exponential
	distributions  with average $\tau_i = \max(\exp(\delta E_i/T),1)$.  
	These random quantities also define, for each spin $i$,
	its flipping time. This is the  time   at which the spin  would flip, 
	were $\delta E_i$  to remain
	constant in the meanwhile. 
	The  simulation iterates the following two steps:  
	\emph{i)}\   the current   time 
	is set to the  earliest  flipping time,  and the 
	corresponding spin,  e.g.\ number   $k$,  is flipped;
	\emph{ii)} \ the   waiting  times of  
	spin $k$  and its nearest neighbors are 
	re-calculated   taking the new  $\delta E_i/T$s
	into account. Adding these waiting times
	to the current time  yields  the updated  flipping times
	of  spin $k$ and its neighbors. All other spins 
	are unaffected by the  move and   need  not
	be updated, as they are engaged in a memory-less  (Poisson)
	 waiting process.   
	 The modification 
	alluded to is that, whenever $T$ changes,  all  $\delta E_i/T$ values 
	are affected and   all   the   
	waiting and  flipping times are 
	re-calculated. 
	
	The data shown are  
	obtained,  for each choice  of physical parameters, from
	$2000$ to $6000$ independent runs. Each run  typically samples 
	$50$ values of $H$ (this  figure varies slightly  according to 
	the parameter settings).  PDFs are thus estimated as 
	the empirical frequencies of  $N=1.5 \cdot 10^5$ data
	points for $150$ bins covering the range of the observations,
	and normalized for convenience to cover a unit area.
	The statistics of the Gaussian part of the data is 
	excellent and does not need further discussion.
	The statistical error on the exponential tail 
	is, unfortunately,  not completely  negligible.   Consider   a 
	data bin whose theoretical 
	probability is $p$, with $p<<1$. The number $n$ of  
	 independent measurements which out of $N>>1$ trials 
	fall  into this bin is well approximated by a  Poisson variable,  with  average
	and variance  both equal to  $pN$. 
	The estimate  $n/N$   for  $p$ has 
	mean $p$ and   standard deviation $(p/N)^{1/2}$.
	The relative error
	on a  data point  with theoretical probability $p$ is the  ratio
	of the standard deviation to the mean 
	and  equals   $(1/pN)^{1/2}$.  The   slow convergence
	with $N$  prevents a perfect   collapse of the tail data. 
	 A  relative  error of   $10$\%, which is still  graphically evident
	in a semi-logarithmic plot,  is obtained for  our range of $N$ 
	for values of $p$ above  $ 10^{-3}$. 
	
	To  skewness (third central moments), which  is used as a  
	measure of the deviation from the equilibrium-like
	Gaussian shape, amplifies the noise in the poorly
	sampled extreme tail of the PDF. To mitigate  the
	effect, we estimated the skewness using 
	 PDFs    curtailed at 
	$10^{-3}$. This  appears to be  sufficiently accurate for  
	comparing  PDF's  describing  different physical situations. 
	  
\section{Summary and discussion}
\label{discussion} 
 Statistical analysis of  mesoscopic noise 
is a powerful probe of 
aging dynamics in a variety of glassy 
systems~\cite{Cipelletti03a,Buisson03,Buisson03a,Sibani04}, as 
reversible equilibrium fluctuations and irreversible drift   
appear as   distinct components of the signal's 
Probability Distribution Function (PDF): the former 
are described by a Gaussian of zero average, and the
latter by an `intermittent'  exponential tail. 
Our investigations utilize the intermittent tail
to describe    the aging behavior of
 intermittent  events, 
or quakes, showing   how their rate  decays with the age and 
how the decay is  modified  by temperature
shifts. A unified
explanation  of both these effects and of pure aging 
behavior~\cite{Sibani03,Sibani04} is achieved  by postulating that
 quakes occur whenever 
a record sized thermal fluctuation crosses the 
extremal barrier established in the course of  
the  preceding thermal history: 
  a  spin  glass is thus claimed to remember  a sequence
of ever growing extremal barriers. This  information 
 is erased
by a positive $T$ shift but preserved by a negative shift.  
The   mechanism  suggests the
presence of  a strong entropic bias
favoring shallow attractors, i.e. ensuring that 
the least stable attractor is selected with
high probability   after each quake. 
 
The  rate of intermittent events  
following a  $T$ shift 
  is thus a   sensitive probe  of  
 the size of the barriers surmounted by an
 aging system before the shift, but 
  carries  no  information
  on e.g.\ the   energy  released through the quakes. The latter 
  crucially affects  the energy density $e(t)$ as a function 
 of the age, whose behavior under $T$ shifts has been 
thoroughly investigated~\cite{Komori00,Berthier02}:
 After temperature shifts,   
 $e(t)$  approaches the isothermal energy density 
 curve obtained at the final temperature. The approach  is  non-monotonic  
 for   positive shifts, a feature   generally called a 
 Kovacs effect. 
As the  energy density is  closely related to the heat transport PDF
a comparison is very instructive.

The effective ages as presently defined
 depend  on  other  dynamical properties 
than those extracted from  $e(t)$, i.e.\ on 
the \emph{rate} of intermittent events rather than the \emph{full shape} of the
energy decay. In particular, they do not 
fully describe the dynamical effect of the $T$ shift, e.g.\ 
 the  shape of  the collapsed PDFs   differs  for  $T$ shifted
and isothermally aged systems. By contrast, in 
energy density studies   the effective age is constructed 
to make the  $T$-shifted and isothermally aged 
curves coincide.  
Furthermore, our definition 
 most directly   relates to   the barriers scaled  \emph{prior}  to  the step and 
 to the  effect of the step itself, unlike 
  the  effective age extracted   from $e(t)$ curves
which most directly gauges   the behavior \emph{after} 
 the thermal step.

In an aging system one expects  the configurational memory  
to vanish on a time scale equal to the system age.  
For a negative $T$ step we expect the memory of the  step to vanish on a time 
scale of the order of $t_w^{eff}>>t_w$, which qualitatively agrees
with the results of refs.~\cite{Komori00,Berthier02} 
This reasoning   does not  apply to a   positive step:
 the  intermittency rates     immediately adjust to those
of an isothermally aged system, while the energy density 
 shoots up  and then  appraoches
the isothermal curve on a time scale which 
is shorter than $t_w$ but differs from zero.  
Clearly, an abrupt change in the barrier structure 
 does not logically imply an equally abrupt 
 change in   energy or configuration, and the data tell us that
the two types of change happen on different scales. 
In summary,  while a naive 
correspondence cannot be established,  a  substantial
 physical similarity   emerges
  when notational and methodological 
differences are properly accounted for. 

The same applies to the  effect of $T$ shifts 
other physical properties, i.e.  the  AC susceptibility $\chi$ 
of  Edwards-Anderson spin-glasses 
in four and three spatial dimensions 
 studied in simulations by 
 Berthier and Bouchaud~\cite{Berthier02} 
and Takayama and Hukushima~\cite{Takayama02}. These authors
parameterize their   results with effective ages 
extracted  in a way similar
to the  energy density analysis just discussed. 
As in our findings,   negative
shift have a double effect:  the apparent rejuvenation 
consisting in a fast decay of   $\chi''$  immediately after 
the shift  and   resembling the evolution right after
the initial quench.  
Secondly,      the curve  
'undershoots' the isothermal data at the final 
temperature, which  makes  the system 
look older. In ref.~\cite{Takayama02}, the relation between
 age and  effective age is nearly a power-law, i.e.\  similar
to Eq.~\ref{effective}.  
The exponent is however $T/T'$, and not 
 our $a(T')/a(T)$, where $a(T)$ is given in  Eq.~\ref{reduced}. 
A positive shift  is there described by an effective
age  shorter  than the  age,  while our  effective age  
simply remains equal to $t_w$. 
This difference  corresponds to 
the difference already discussed in connection
with  the $e(t)$ data, but 
the    asymmetry between positive and negative shift  
  found  in ref.~\cite{Takayama02}  is similar to our findings.

Comparing simulation studies with experimental results,
as  we shall do next,  
 requires an element of faith,   particularly so   for dynamical 
effects: simulation algorithms such as
the Metropolis algorithm and the WTM presently 
utilized are designed  to sample   equilibrium
distributions, but the time scales
of  e.g.  partial equilibrations 
are  algorithm dependent and cannot be meaningfully 
connected to experimental 
time scales. A large empirical evidence nevertheless points to the  
 relevance  for  dynamical phenomena 
of the  information gained from simulations of glassy  
systems, perhaps unsurprisingly,   given that the  dynamical 
features experimentally defining glassy behavior are similar 
across a wide range of microscopically very different systems,
and  that aging dynamics is, broadly speaking, controlled by 
time scales ratios, rather than   magnitudes. 

The temperature
and age dependence studies of  the imaginary part of the AC  
susceptibility $\chi''$~\cite{Jonason98},
show  how the memory of configurations 
explored  at a certain temperature survives cooling and is retrieved by
re-heating. This  information is  not  directly available 
from  heat transfer intermittency  data. However, 
the related property that the   memory  is  erased by 
a sufficiently strong overheating~\cite{Jonason00}
is qualitatively similar to  the asymmetry between positive
and negative step already discussed.  

The relationship  between effective and actual age
 under positive and negative temperature
shifts is  investigated experimentally
in ref.~\cite{Jonsson02}. 
For sufficiently large $T$ shifts, a lack of reversibility of the $T$ step, 
called  a `non-accumulative  aging effect'  is present.
This is   qualitatively similar  to 
our findings, where the  shifts applied are of the order of   
$5$ \% of $T_g$, 
i.e.   large  compared to those 
 used in ref.~\cite{Jonsson02}.  
We can thus  confirm  our 
 conclusion that the  physical insight 
obtained  from the intermittency
analysis broadly concurs with the 
results of other types of investigation.

As intermittent  heat transfer  
  says    only little about  configuration
changes,   information regarding   
the high temperature sensitivity of the thermal
correlation length, so called 'chaos'~\cite{Bray87}, which is
often linked to  rejuvenation effects, can only be gained
indirectly. 
 We note however  that   since  memory and rejuvenation 
effects  appear  in systems  of very modest size,  
  properties of  large scale thermal
 objects, e.g.\ droplet excitations 
out of a ground state~\cite{Hartmann04} are largely immaterial 
for the present discussion. 

As  recently shown~\cite{Dall03}, geometrical properties
of the  energy landscape
of a spin-glass depend in a systematic way on the 
extremal energy barrier crossed.
More generally, a   linear relationship between the average 
energy barrier crossed  and  the configurational
changes induced by the crossing has been measured experimentally~\cite{Lederman91}
and    numerically,  in different types of spin-glass
models~\cite{Vertechi89,Dall03} and using  different 
exploration techniques~\cite{Boettcher04}.
Due to the strong connection between barriers overcome
and configuration changes we can expect 
that erasing the memory of the extremal barrier 
would also lead to some, more gradual,  loss of configuration memory.

A linear relationship between 
barriers and Hamming distances 
is included  in  descriptions of spin-glass
dynamics as a master equation on 
a  set of hierarchically 
organized states~\cite{Sibani89,Schultze91,Sibani91,Hoffmann97,Joh96}.  
The tree   models   used in these descriptions are
energy landscape descriptions of 
local regions of a complex system, and  are able
to reproduce salient  features
of aging, including $t/t_w$ scaling, and the effects
of temperature shifts based on thermal hopping.
rejuvenation effects
have also been found~\cite{Hoffmann97} in the so-called LS tree~\cite{Sibani91},
based on the idea that a kinetic bias exist favoring 
shallow attractors. 
The record induced dynamics scenario   presently advocated also  requires 
 the presence of a hierarchy of inequivalent dynamical attractors.
However, unlike the case of a static tree model, 
 these attractors are not given in advance, but 
are dynamically selected by  record energy fluctuations. 

Unlike other methods,
the  intermittency  probe of aging dynamics has a \emph{direct}
 analytical interpretation, 
and, at least for spin-glasses, appears to offer a different
   age  and temperature window  where it can conveniently 
be applied.   I.e.  intermittency rates 
 can be obtained  for all ages, whether  isothermal evolution is fast
 or slow, but are most conspicuous  at low temperatures (below $T\approx 0.6$
 for the present   system) where  
 Gaussian fluctuations are small~\cite{Sibani04}.  
 
Secondly, since  intermittent events entail large 
configurational rearrangements, they  expectedly 
 influence  in  similar  ways   different  physical properties.  
E.g.\ studies of 
  the electrical voltage fluctuations  
in Laponite, a gel,  and in a polycarbonate glass~\cite{Buisson04} 
show  that  the intensity of the intermittent signal
decreases while the system ages (as in our fig.~\ref{introplot}).
In spin-glasses, where the spin contribution to the
energy is not easily separated out, one could  study the intermittency property of 
the magnetic noise or, in metallic spin-glasses,
  the electrical fluctuations associated
with spin flips ~\cite{Feng87}. 
The method presented
  can in principle be applied  to analyze intermittency
rates of  any  relevant  and experimentally available quantity.  
Measuring the rate
of intermittent events as a function of the  age 
would be extremely  interesting from a theoretical 
point of view, but  also extremely challenging. 
Some  changes of the PDF's cause  by  temperature shifts,  
 see e.g.\ Fig.~\ref{introplot},    
 might be observable   even in the lack of  precise rate measurements.

    \section{Acknowledgments}   Financial support from
    the Danish SNF and from EPSRC  is gratefully acknowledged. 
    \bibliographystyle{unsrt}

\bibliography{SD-meld}

\end{document}